\documentclass[conference]{IEEEtran}

\pdfoutput=1

\usepackage{epsfig}
\usepackage{amsmath}
\usepackage{amssymb}
\usepackage{color}
\usepackage{url}

\newlength{\figurewidth}
\newlength{\smallfigurewidth}

\def\bfx{{ \bf x  }}

\def\defeq{{\stackrel{\Delta}{=}}}
\def\calX{{\mathcal{X}}}

\newtheorem{theorem}{Theorem}[section]

\newtheorem{proposition}[theorem]{Proposition}

\DeclareMathOperator*{\argmin}{arg\!\min}

\begin{document}


\title
{\LARGE
\textbf{Minimum Conditional Description Length\\ Estimation for Markov Random Fields}
}

\author{%
{\small\begin{minipage}{\linewidth}\begin{center}
\begin{tabular}{ccc}
{\large Matthew G. Reyes}$^{\ast}$ & \hspace{20mm} & {\large David L. Neuhoff}$^{\dag}$\\
$^{\ast}$self-employed & \hspace*{0.5in} & $^{\dag}$EECS Dept., University of Michigan \\
\url{mgreyes@umich.edu} && \url{neuhoff@umich.edu}
\end{tabular}
\end{center}\end{minipage}}
}

\maketitle
\thispagestyle{empty}

\begin{abstract}
In this paper we discuss a method, which we call Minimum Conditional Description Length (MCDL), for estimating the parameters of a subset of sites within a Markov random field. We assume that the edges are known for the entire graph $G=(V,E)$. Then, for a subset $U\subset V$, we estimate the parameters for nodes and edges in $U$ as well as for edges incident to a node in $U$, by finding the exponential parameter for that subset that yields the best compression conditioned on the values on the boundary $\partial U$. Our estimate is derived from a temporally stationary sequence of observations on the set $U$. We discuss how this method can also be applied to estimate a spatially invariant parameter from a single configuration, and in so doing, derive the Maximum Pseudo-Likelihood (MPL) estimate.
\end{abstract}

\section{Introduction}\label{sec:introduction}
A Markov random field (MRF), also referred to as a Gibbs distribution, is a probability distribution on the colorings of an undirected graph $G=(V,E)$, where the nodes\footnote{We use the terms \emph{nodes} and \emph{sites} interchangeably.} in $V$ are the random variable indices and the edges in $E$ represent direct dependencies between the random variables \cite{wain:03b}. One of the primary research areas for MRFs is the problem of model selection or parameter estimation, where the objective may either be to determine the parameters for known edges \cite{besag1974}, determine the edges of the graph \cite{csiszar2006}, or jointly find the edges and the parameters for those edges \cite{pietra1997}. Markov fields are a natural class of models for many types of data, including images and social networks. In images, it is natural to assume a set of edges, for instance, those connecting the 4 or 8 nearest neighbors. And for social networks, neighbor relations are known. With these two applications in mind, this paper focuses on the first model selection problem, that of determining the parameters on known edges.

A family of MRFs is specified by a vector statistic $t=(t_i, i \in V;  t_{i,j}, \{i,j\} \in E)$ defined on the site values at individual nodes and the endpoints of the edges $E$ of the graph.\footnote{Properly, this is a {\em pairwise} MRF. Generalizations to other MRFs are straightforward.} A particular MRF is indexed by an exponential parameter vector $\theta$ that scales the corresponding components of $t$ in the probability of a configuration $\bfx$, which is given by
\begin{eqnarray}
    p(\bfx;\theta)
                & = &
                \exp\{\langle\theta,t(\bfx)\rangle-\Phi(\theta)\},\label{eq:mrf_1}
\end{eqnarray}
\noindent where  $\langle,\rangle$ denotes inner product and $\Phi(\theta)$ is the {\em log-partition function}.

In the {\em model selection} problem considered in this paper, the set of edges $E$ is known, as well as the statistic $t$, and we have to determine the exponential parameter $\theta$ that weights the corresponding components of the statistic for nodes and edges. Generally, estimation is performed from a temporal sequence of observations $\bfx^{1:n}\defeq\bfx^{(1)},\ldots,\bfx^{(n)}$, from which an estimate $\hat\theta^n$ is obtained. While it is often assumed that the $\bfx^{(i)}$ are independent to simplify analysis, in fact it is sufficient to assume that $\bfx^{(1)},\bfx^{(2)},\ldots$ is stationary, which is what we assume in this paper.

A popular criterion for estimating a parameter within a family of candidate models is {\em Maximum Likelihood} (ML), which seeks the parameter $\hat\theta^n$ which maximizes the probability $p(\bfx^{1:n};\tilde\theta)$ of the observed data over all parameter vectors $\tilde\theta$ indexing probability distributions within the specified class of probability distributions. For Markov fields, the ML criterion reduces to finding the exponential parameter $\tilde\theta$ such that the expected statistic $\tilde\mu\defeq\mu(\tilde\theta)\defeq\mathbb{E}_{\tilde\theta}[t(X)]$ under the MRF induced by $\tilde\theta$, referred to as the {\em moment} of the MRF, equals the {\em empirical moment} $\hat\mu^n$ of $\bfx^{1:n}$, which is the average value  $\frac{1}{n}\sum_{i=1}^nt(\bfx^{(i)})$ of the statistic from the $n$ observations \cite{geyer1992}. For a tractable graph, such as a tree or one that can be clustered into a tree with only moderate numbers of nodes per cluster, the moments can be exactly and efficiently determined with Belief Propagation (BP), an iterative message passing algorithm. Thus, one can compute moments $\{\tilde\mu\}$ for a set of candidates $\{\tilde\theta\}$ and choose the one whose moment $\tilde\mu$ most closely matches the observed empirical moment $\hat\mu^n$. For a general graph, however, BP is intractable and thus the moment $\tilde\mu$ cannot be computed exactly. This intractability can circumvented by approximating the moment $\tilde\mu$, with either an approximate variant of BP \cite{wain:03b}, or by sampling the MRF's corresponding to candidate $\tilde\theta$, e.g. with Gibbs sampling \cite{gema:84}, \cite{geyer1992}, \cite{geyer1994}, and selecting the $\tilde\theta$ whose empirical moment $\hat{\tilde\mu}$ most closely matches that of the observed data.

An alternative method for making parameter estimation in MRFs tractable is Maximum {\em Pseudo-Likelihood} \cite{besag1977}, which defines a different objective function that is tractable and hence can be solved exactly. Maximum Pseudo Likelihood (MPL) is based on the concept of a {\em Coding Method}, introduced by Besag \cite{besag1974}. Assuming a translation invariant statistic $t$ as well as a translation invariant parameter $\theta$, such that each site had the same conditional distribution conditioned on its {\em neighbors}, those sites connected to it by an edge, one chooses a subset $V_1\subset V$ of sites such that no two sites in $V_1$ are neighbors in $G$. By the Markov property, the sites in $V_1$ are conditionally independent of one another conditioned on the sites in $V\setminus V_1$, permitting their conditional distribution to be expressed as a product of single-site conditional probabilities. Thus, by conditioning on $\bfx_{V\setminus V_1}$, one can estimate $\theta$ through an analytically tractable objective function. MPL extends this idea by finding the parameter $\hat\theta^{MPL}$ that maximizes the {\em pseudo}-likelihood function
\begin{eqnarray}
    \mbox{PL}(\bfx;\tilde\theta) & = & \prod_{j=1}^{|V|}p(\bfx_j | \bfx_{V\setminus j};\tilde\theta) \nonumber
\end{eqnarray}
over candidate parameters $\tilde\theta$, or equivalently, the pseudo-log-likelihood function
\begin{eqnarray}
    \log\mbox{PL}(\bfx;\tilde\theta) & = & \sum\limits_{j=1}^{|V|}\log p(\bfx_j | \bfx_{V\setminus j};\tilde\theta), \nonumber
\end{eqnarray}
again assuming translation invariance, or spatial homogeneity, of $t$ and $\theta$. Again by the Markov property, these conditional probabilities simplify as conditional probabilities given the neighbors of each node. Much research has been done on MPL, and consistency of the MPL estimate $\hat\theta^{MPL}$ has been shown \cite{gidas}, \cite{comets1992}. An interpretation of MPL is that it finds the parameter $\hat\theta^{MPL}$ such that the induced conditional distributions of individual nodes best match the empirical conditional distributions of individual nodes.

The parameter estimation method proposed in the present paper, which we call Minimum Conditional Description Length (MCDL), can be understood as a generalization of Maximum Pseudo-Likelihood. Whereas the MPL method estimates a translation invariant parameter through observations $\bfx_{\bar U_1},\ldots,\bfx_{\bar U_n}$ of $n=|V|$ statistically identical subsets within a single observation $\bfx$, we propose MCDL as a method for estimating the parameter $\theta_{\bar U}$ within a single subset $\bar U$ from a sequence of observations $\bfx_{\bar U}^{(1)},\ldots,\bfx_{\bar U}^{(n)}$ on $\bar U$, where $\partial U$ is the boundary or neighborhood of $U$ and $\bar U = U\cup\partial U$ is the {\em closure} of $U$. We do not assume spatial homogeneity (translation invariance) of $\theta$ within $G$, but we do require temporal stationarity of $\bfx_{\bar U}^{(1)},\ldots,\bfx_{\bar U}^{(n)}$. Moreover, while in MPL the subsets $U_j$ are single sites, here the only restriction we place on a subset $U$ is that the {\em subgraph induced by} $U$, consisting of nodes and edges of $G$ contained in $U$, be tractable with respect to BP.

The {\em Minimum Description Length} (MDL) \cite{rissanen1978} principle states essentially that the best model is that one that provides the best compression of the data. Since Markov fields are defined in terms of their conditional distributions, and since conditioning on the boundary of a subset renders the subfield within the subset conditionally independent of the subfield outside of the closure of the subset, MCDL is a natural extension of this for efficiently estimating the parameters $\theta_{\bar U}$ inducing the conditional distribution of $X_U$ given $X_{\partial U}$. If subset $U$ is tractable for BP, we can compute the conditional probability
\begin{eqnarray}
    p(\bfx^{(i)}_U | \bfx^{(i)}_{\partial U};\tilde\theta_{\bar U}) \nonumber
\end{eqnarray}
of a configuration $\bfx^{(i)}_U$ given the configuration $\bfx^{(i)}_{\partial U}$ on its boundary. Then, given a temporal sequence of configurations $\bfx_{\bar U}^{1:n}=(\bfx^{(1)}_{\bar U},\bfx^{(2)}_{\bar U},\ldots,\bfx^{(n)}_{\bar U})$ on the closure $\bar U$, we seek the parameter $\hat\theta_U=\hat\theta^n_{\bar U}$ that causes the conditional distribution of $\bfx_U$ given $\bfx_{\partial U}$ within the MRF modeled by $\hat\theta_{\bar U}$ to best approximate the empirical conditional distribution of the $(\bfx^{(i)}_U:1\leq i\leq n)$ conditioned on the corresponding values $(\bfx^{(i)}_{\partial U}:1\leq i\leq n)$ on the boundary. Thus for different candidate parameters $\tilde\theta_{\bar U}$ we compute the temporal average of the negative log likelihood
\begin{eqnarray}
       H^n_{\bar U}(\tilde\theta_{\bar U}) & = & \frac{1}{n}\sum\limits_{i=1}^n-\log p(\bfx^{(i)}_U | \bfx^{(i)}_{\partial U};\tilde\theta_{\bar U}) \label{eq:cross}
\end{eqnarray}
\noindent and select the $\tilde\theta_{\bar U}$ that minimizes $H^n_{\bar U}(\tilde\theta_{\bar U})$. It is important to note that while $\tilde\theta_{\bar U}$ is properly the parameters for all nodes and edges within the closure $\bar U$ of $U$, the conditional distribution $p(X_U|\bfx_{\partial U};\tilde\theta_{\bar U})$ of $X_U$ given $\bfx_{\partial U}$ depends only on the parameters for nodes and edges within $U$ and for those edges connecting $U$ to $\partial U$. It is in this more restricted sense that we use $\tilde\theta_{\bar U}$ throughout this paper.

This average negative log-likelihood can be interpreted as an empirical cross entropy between the true conditional distribution induced by $\theta_{\bar U}$ and the candidate parameter $\tilde\theta_{\bar U}$. Note that if $\bfx^{(1)},\ldots,\bfx^{(n)}$ were independent, this would be the negative log likelihood and this method would produce the ML estimate for $\theta_{\bar U}$. With an optimal encoder, for example Arithmetic Coding (AC) \cite{whitten87}, for each $i$ the number of bits produced in encoding $\bfx_U^{(i)}$ conditioned on $\bfx_{\partial U}^{(i)}$ will be within 1 or 2 bits of $-\log p(\bfx_U^{(i)} | \bfx_{\partial U}^{(i)};\tilde\theta_{\bar U})$. In other words, deriving the estimate $\hat\theta^n_{\bar U}$
%
%
as the parameter subvector that minimizes cross-entropy is essentially equivalent to estimating $\theta_{\bar U}$ as the parameter that minimizes coding rate when conditionally coding $X_U$ given $X_{\partial U}$ with conditional coding distribution induced by $\tilde\theta_{\bar U}$.
Indeed, it is straightforward to show that in the limit as the number of temporal samples $n$ tends to infinity, the empirical average $\frac{1}{n}\sum_{i=1}^n-\log p(\bfx_U^{(i)} | \bfx^{(i)}_{\partial U};\tilde\theta_{\bar U})$ converges to
\begin{equation}
    H(X_U | X_{\partial U};\theta) + D(p(X_U | X_{\partial U};\theta_{\bar U}) || p(X_U | X_{\partial U};\tilde\theta_{\bar U})) \nonumber
\end{equation}
for a given candidate parameter $\tilde\theta$.

Ultimately, this method would be applied to different subsets $U_1,\ldots,U_k$, yielding estimates $\hat\theta_{\bar U_1},\ldots,\hat\theta_{\bar U_k}$ for the conditional distributions of $X_{U_1},\ldots,X_{U_k}$ given their respective boundaries. In order to produce an estimate $\hat\theta$ of the full parameter vector, we would need a way to enforce consistency of the $\hat\theta_{\bar U_1},\ldots,\hat\theta_{\bar U_k}$ on nodes and edges contained in multiple $\bar U_j$. At the moment we focus on estimating $\theta_{\bar U}$ for a single subset $U$.

%
%

To reiterate, one way in which MCDL differs from MPL is in the stationarity or homogeneity assumptions used to obtain the statistics for estimation. The setting in which MPL is generally applied assumes a translation invariant exponential parameter $\theta$ on a regular graph, in particular where the set of sites $V$ form a lattice, and where an estimate of the global parameter $\theta$ is obtained from a single observation $\bfx$ on $V$. We do not require spatial homogeneity of the parameter, though we do require temporal stationarity and estimate the parameter for a single subset from a temporal sequence of observations on that subset. In other words, whereas we are proposing to estimate the parameters $\theta_{\bar U}$ through $n$ observations $\bfx_{\bar U}^{(1)},\ldots,\bfx_{\bar U}^{(n)}$ on given subset $U$ and its boundary, the MPL method estimates a translation invariant parameter $\theta$ through observations $\bfx_{\bar U_1},\ldots,\bfx_{\bar U_n}$ on $n$ statistically identical subsets $U_1,\ldots,U_n$ and their boundaries within a single observation $\bfx$.


The proposed MCDL algorithm also differs from MPL in that it allows larger subsets $U$ rather than single sites, and more conceptually, in the formulation of the objective function. We now digress for a moment to think about MPL in the context of these other two differences. A common remark in the literature is that while the pseudo-likelihood function is tractable it is viewed as an approximation to the (chain rule decomposition of) the true likelihood function $p(\bfx;\tilde\theta)$ of the observed data. However, in the translation invariant setting of MPL analysis, rather than attempt to approximate the likelihood function, instead consider the MCDL objective function, the cross entropy
\begin{eqnarray}
    H^n(\tilde\theta) & \defeq & \frac{1}{n}\sum\limits_{i=1}^n-\log p(\bfx_{i} | \bfx_{\partial i};\tilde\theta) \label{eq:cross_mpl}
\end{eqnarray}
between the empirical conditional distributions of single sites and the single site conditional distributions induced by a candidate parameter $\tilde\theta$. Mathematically, we have the same objective function for a candidate parameter $\tilde\theta$. However, viewed through the lens of MCDL, this function now yields the parameter that achieves minimal conditional description length of a site conditioned on its neighbors, without recourse to anything `pseudo' or approximate. Indeed, in the limit of a large lattice of sites $V$, Equation (\ref{eq:cross_mpl}) above tends to
\begin{eqnarray}
    H^-(X;\theta) + D(p(X_0 | X_{\partial 0};\theta) || p(X_0 | X_{\partial 0};\tilde\theta)), & & \label{eq:coding_MPL}
\end{eqnarray}
where $H^-(X;\theta)$ is the {\em erasure entropy} \cite{verdu2008}, given by
\begin{eqnarray}
    H^-(X;\theta) = H(X_0 | X_{\partial 0};\theta), \nonumber
\end{eqnarray}
which is the information lost if $X_0$ is erased from $X$, or in other words, the minimal amount of information needed to describe it conditioned on the values of its neighbors. It should be noted that (\ref{eq:coding_MPL}) is not the number of bits from a lossless code of $X$, as clearly $H^-(X) < H(X;\theta)$. Nonetheless, through the MCDL paradigm, the MPL estimate can be interpreted as minimizing the empirical coding rate of $\{\bfx_{U_i}\}$ conditioned on the values $\{\bfx_{\partial U_i}\}$ rather than as an approximation of the likelihood function. Since Markov/Gibbs fields are specified in terms of their local characteristics, i.e., their conditional distributions, it makes perfect sense that MPL would yield a consistent estimate of $\theta$.

Moreover, casting MPL as a conditional description length problem, one can generalize from considering conditional distributions of single nodes to considering conditional distributions of larger subsets $U_i$. Then for an MRF induced by a translation invariant parameter $\theta$, the objective function to be minimized is now
\begin{eqnarray}
    \frac{1}{n}\sum\limits_{i=1}^n-\log p(\bfx_{U_i} | \bfx_{\partial U_i};\tilde\theta). && \nonumber
\end{eqnarray}
As opposed to subsets $U_i$ of size 1, using larger subsets will reduced the number of samples $n$, so in that sense could potentially have an adverse affect on convergence and therefore the accuracy of $\hat\theta^n$. On the other hand, as the subsets $U_i$ become larger, the effect of conditioning is reduced relative to the inter-site interactions within the $U_i$ and as a result the local characteristics within a $U_i$ conditioned on its boundary $\partial U_i$ will more closely approximate the local characteristics of the full distribution. In other words, it is worth examining the tradeoffs involved in using larger subsets. Moreover, considering larger subsets $U_i$ allows for greater flexibility in the invariance required for this method to provide good estimates. For example, instead of requiring site invariance of the statistic and parameter, one could simply assume row invariance of the statistic and parameter in which case the subsets $U_i$ would be different rows of the lattice.

We now return to MCDL and consider the task of showing that the estimate $\hat\theta^n_{\bar U}$ of $\theta_{\bar U}$ is consistent, that is, that $\hat\theta^n_{\bar U} \rightarrow \theta_{\bar U}$ as $n\rightarrow\infty$. A reasonable course of action would be to mimic as closely as possible the proofs of consistency of the MPL estimate \cite{gidas}, \cite{comets1992}. The only difference it seems is that in the MPL regime, the $X_{U_1},\ldots,X_{U_n}$ are independent conditioned on their respective boundaries, whereas in our case the $X^{(1)}_U,\ldots,X^{(n)}_U$ are not independent conditioned on the boundaries. Both problems have the same objective function, however, so it remains to be seen just how much tweaking is required to extend the MPL results to the present paradigm.

In the rest of this paper, Section \ref{sec:background} provides background on MRFs. Section \ref{sec:bel} discusses the use of BP in lossless coding, Section \ref{sec:selection} presents our algorithm for estimating the parameter within a subset, and Section \ref{sec:example} discusses an example where we apply MCDL to both temporally stationary observations on a single subset as well as spatially observations on multiple subsets of a single configuration.

\vspace{2mm}

\section{Graphs and Markov Random Fields}\label{sec:background}

At each site $i\in V$ there is random variable $X_i$ assuming values in alphabet $\mathcal{X}_i$. For a given configuration $\bfx = \{x_i: i\in V\}$ , the function $t_{ij}:\calX_{i}\times\calX_j\longrightarrow\mathbb{R}$ determines the contribution of the pair $(x_i,x_j)$ to the probability of $\bfx$, and similarly for $t_i:\calX_i\longrightarrow\mathbb{R}$. We say that $X=(X_i,i\in V)$ is an MRF based on $t$. The entire family of MRFs based on $t$ is generated by introducing an exponential parameter $\theta=(\theta_i, i \in V; \theta_{ij}, \{i,j\} \in E)$ where for each node $i$, and neighbor $j\in\partial i$, $\theta_i$ and $\theta_{ij}$ scale the
sensitivity of the distribution $p(\bfx)=p(\bfx;\theta)$ to the functions $t_i$ and $t_{ij}$, respectively.

The conditional probability of a configuration $\bfx_U$ on subset $U\subset V$ given the values $\bfx_{W}$ on another subset $W\subset V$ is denoted $p(\bfx_U| \bfx_{W};\theta)$. It is straightforward to check that $p(\bfx_U| \bfx_{\partial U};\theta) = p(\bfx_U | \bfx_{V\setminus U};\theta)$ for all $U$, $\bfx_U$, and $\bfx_{\partial U}$. This is the {\em Markov Property}. The conditional distributions of random subfield $X_U$ given a specific configuration $\bfx_{\partial U}$, or on the random subfield $X_{\partial U}$, are denoted $p(X_U| \bfx_{\partial U};\theta)$ and $p(X_U | X_{\partial U};\theta)$, respectively. Likewise $H(X_U | \bfx_{\partial U};\theta)$ and $H(X_U | X_{\partial U};\tilde\theta)$ are the respective conditional entropies of $X_U$ given a specific configuration $\bfx_{\partial U}$ or the random subfield $X_{\partial U}$.

It is straightforward to show the following.

\vspace{2mm}

\begin{proposition}
    \begin{eqnarray}
        p(\bfx_U | \bfx_{\partial U};\tilde\theta_{\bar U}) & = & \exp\{\langle t_{\bar U}(\bfx_U,\bfx_{\partial U}),\tilde\theta_{\bar U}\rangle - \Phi_{U | \bfx_{\partial U}}(\tilde\theta_{\bar U}\},\nonumber
    \end{eqnarray}
    \noindent where
    \begin{eqnarray}
        \Phi_{U | \bfx_{\partial U}}(\tilde\theta_{\bar U}) & = & \log\left[\sum\limits_{\bfx'_U}\exp\{\langle t_{\bar U}(\bfx'_U,\bfx_{\partial U}),\tilde\theta_{\bar U}\rangle\}\right] \nonumber
    \end{eqnarray}
    \noindent is the log partition function for the conditional distribution of $X_U$ with boundary condition $\bfx_{\partial U}$. Note that the statistic $t_{\bar U}(\bfx'_U,\bfx_{\partial U})$ includes all components of $t$ at least one argument of which is contained in $U$. Thus $p(\bfx_U | \bfx_{\partial U};\tilde\theta_{\bar U})$ does not depend on $\tilde\theta_{\partial U}$.
\end{proposition}

\vspace{2mm}

\section{Belief Propagation and Minimum Description}\label{sec:bel}

In general, ones uses Belief Propagation (BP) \cite{wain:03b} to compute $p(\bfx;\theta)$ for a configuration $\bfx$. Since the inner product $\langle t(\bfx),\theta \rangle$ can be computed directly, BP is used to (indirectly) compute the normalizing constant, the log-partition function $\Phi(\theta)$. If $G$ has no cycles, then $p(\bfx;\theta)$ can be computed with complexity linear in the number of nodes in $V$. If $G$ has cycles, one can compute $p(\bfx;\theta)$ by grouping subsets of $V$ into supernodes such that the new graph is acyclic \cite{wain:03b}. In this case, complexity is exponential in the size of the largest supernode. A graph is said to be {\em tractable} if either $G$ has no cycles or if $G$ can be clustered into an acyclic graph where the size of the largest supernode is moderate, for example no more than 10. A subset $U$ is said to be tractable if the subgraph induced by $U$ is tractable. For tractable subset $U$, $p(\bfx_U | \bfx_{\partial U};\theta)$ can be computed for given configurations $\bfx_U$ and $\bfx_{\partial U}$. Specifically, the conditional probability distribution $p(X_U | \bfx_{\partial U};\theta)$ of $X_U$ given the configuration $\bfx_U$ on $\partial U$ can be computed exactly and efficiently.

For the purposes of this paper it suffices to say that lossless compression with an {\em optimal encoder} involves computation of a {\em coding distribution}. For a tractable subset $U$, if configuration $\bfx_{U}$ is encoded conditioned on $\bfx_{\partial U}$ using coding distribution $p(X_U|\bfx_{\partial U};\tilde\theta_{\bar U})$, then the average number of bits produced is
\begin{eqnarray}
    \lefteqn{H(X_U | X_{\partial U} ;\theta_{\bar U} || X_U | X_{\partial U};\tilde\theta_{\bar U}) \defeq } \nonumber\\
        & & \!\!\!\! H(X_U | X_{\partial U};\theta_{\bar U}) + D(p(X_U | X_{\partial U};\theta_{\bar U}) || p(X_U | X_{\partial U} ;\tilde\theta_{\bar U}))\nonumber
\end{eqnarray}
where $D(p(X_U | X_{\partial U};\theta_{\bar U}) || p(X_U | X_{\partial U};\tilde\theta_{\bar U}))$ is the {\em divergence} between $p(X_U | X_{\partial U};\theta_{\bar U})$ and $p(X_U | X_{\partial U};\tilde\theta_{\bar U})$ and is the {\em redundancy} in the code \cite{cove:05}. Clearly, then, the true parameter $\theta_{\bar U}$ eliminates the redundancy and achieves the minimal conditional description length. In \cite{reyes2010}, Arithmetic Coding (AC) was proposed as the optimal encoder and details on the use of AC in the encoding of an MRF are given in \cite{reyes2009t}, \cite{reyes2010}, and \cite{reyes2011}.

\vspace{2mm}

\section{Model Selection by Conditioning}\label{sec:selection}

We now discuss the MCDL method for estimating the parameters $\theta_{\bar U}$ of a subset $\bar U$. For tractable subset $U$, recall that we can exactly compute the probabilities $\{p(\bfx_U^{(i)} | \bfx_{\partial U}^{(i)};\tilde\theta_{\bar U})\}$ since $U$ is chosen to be tractable with respect to Belief Propagation. If $\tilde\theta_{\bar U}$ is the parameter for the conditional distribution used to encode $\bfx^{(i)}_U$ given $\bfx^{(i)}_{\partial U}$, then the codeword length for $\bfx^{(i)}_U$ conditioned on $\bfx^{(i)}_{\partial U}$ is approximately
\begin{eqnarray}
    -\log p(\bfx^{(i)}_U\mid \bfx^{(i)}_{\partial U};\tilde\theta_{\bar U}) && \nonumber
\end{eqnarray}

To form the estimate $\hat\theta^n_{\bar U}$ from observations $(\bfx^{(1)}_U,\bfx^{(1)}_{\partial U}),\ldots,(\bfx^{(n)}_U,\bfx^{(n)}_{\partial U})$, we use BP to compute the empirical cross entropy given in (\ref{eq:cross})
for a candidate parameter $\tilde\theta_{\bar U}$ and then seek to minimize $H^n_{\bar U}(\tilde\theta_{\bar U})$ over $\tilde\theta_{\bar U}$. The following is straightforward to show.

\vspace{2mm}

\begin{proposition}
    \begin{eqnarray}
        H^n_{\bar U}(\tilde\theta_{\bar U}) & = & \frac{1}{n}\sum\limits_{i=1}^n\Phi_{U|\bfx^{(i)}_{\partial U}}(\tilde\theta_{\bar U}) - \langle \hat\mu^n_{\bar U},\tilde\theta_{\bar U}\rangle \nonumber \\
    \nabla H_U^n(\tilde\theta_{\bar U}) & = & \frac{1}{n}\sum\limits_{i=1}^n\tilde\mu_{\bar U|\bfx^{(i)}_{\partial U}} - \hat\mu^n_{\bar U}\nonumber
    \end{eqnarray}
    where $\hat\mu^n_{\bar U} = \frac{1}{n}\sum\limits_{i=1}^nt_{\bar U}(\bfx^{(i)}_U,\bfx^{(i)}_{\partial U})$ is the empirical moment for the subset $\bar U$ given $\{\bfx^{(i)}_{\partial U}\}$ and $\tilde\mu_{\bar U | \bfx^{(i)}_{\partial U}}$ is the conditional moment for $p(X_U| \bfx^{(i)}_{\partial U};\tilde\theta_{\bar U})$.
\end{proposition}

\vspace{2mm}

It is well-known that $\Phi_{U | \bfx^{(i)}_{\partial U}}(\tilde\theta_{\bar U})$ is convex in $\tilde\theta_{\bar U}$ for each $i$, and as such so is $H^n_{\bar U}(\tilde\theta_{\bar U})$. If the components of $t$ are affinely independent, then $H_{\bar U}^n(\tilde\theta_{\bar U})$ is strictly convex and thus has a unique minimum. Note that we are able to compute $\tilde\mu_{\bar U | \bfx^{(i)}_{\partial U}}$ with BP because $U$ was chosen to be tractable. We can therefore apply a gradient descent algorithm to minimize $H^n_{\bar U}(\tilde\theta_{\bar U})$ and obtain the estimate
\begin{eqnarray}
    \hat\theta^n_{\bar U} & = & \argmin\limits_{\tilde\theta_{\bar U}}H_{\bar U}^n(\tilde\theta_{\bar U})\nonumber
\end{eqnarray}

The MCDL algorithm for estimating the parameter subvector $\theta_{\bar U}$ within a subset $U$ can be summarized as follows. Given $(\bfx^{(1)}_U,\bfx^{(1)}_{\partial U}),\ldots,(\bfx^{(n)}_U,\bfx^{(n)}_{\partial U})$, we initially compute the empirical moment $\hat\mu^n_{\bar U} = \frac{1}{n}\sum_{i=1}^nt_{\bar U}(\bfx^{(i)}_U,\bfx^{(i)}_{\partial U})$. Then for a candidate parameter $\tilde\theta_{\bar U}$, we compute $H^n_{\bar U}(\tilde\theta_{\bar U})$ and the $\{\tilde\mu_{\bar U | \bfx^{(i)}_{\partial U}}\}$ using BP. We then compute the gradient $\nabla H^n_{\bar U}(\tilde\theta_{\bar U}) = \frac{1}{n}\sum_{i=1}^n\tilde\mu_{\bar U | \bfx^{(i)}_{\partial U}} - \hat\mu^n_{\bar U}$. Using a standard search, we select a new $\tilde\theta_{\bar U}$, and continue this process until a desired threshold for the norm of $\nabla H^n_{\bar U}(\tilde\theta_{\bar U})$ is attained \cite{boyd2004}.

\vspace{2mm}

\begin{figure*}
    \centerline{    \hbox{
    \hspace{0in}
    \includegraphics[scale = .4]{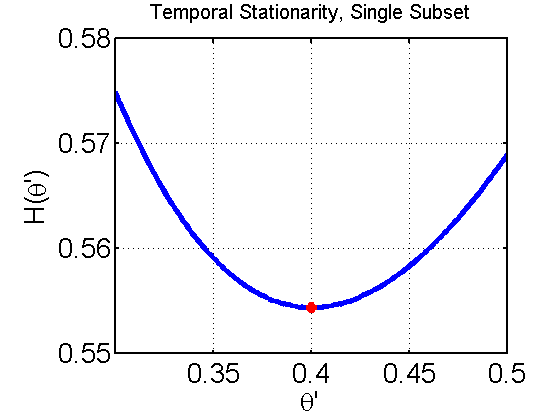}
    \hspace{1.5in}
    \includegraphics[scale = .4]{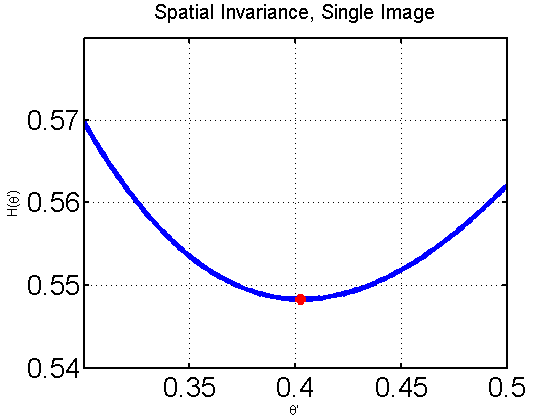}
    }   }
    \hbox{\hspace{1.65in} (a) \hspace{3.65in} (b)}
    \caption{True parameter is $\theta=.4$. Minimizing $\theta'$ is indicated in red in each case. Plot of empirical cross entropy for (a) temporally stationary sequence on a single subset, and (b) spatially invariant parameter on multiple subsets.}
    \label{fig:cross}
\end{figure*}

\section{Example: Homogeneous Ising Model}\label{sec:example}

We experimented with a (spatially) homogeneous Ising model, with edge parameter $\theta_{ij} = .4$ and node parameter $\theta_i = 0$ on a $200\times 200$ square grid of sites, where each interior site is connected to its four nearest neighbors. The results are show in Figure \ref{fig:cross}. In (a), we consider a single subset $U$ that is the middle row of the grid. The boundary $\partial U$ consists of the row above and the row below. We generated a sequence of $n=198$ configurations on $G$ and computed
\begin{eqnarray}
    -\frac{1}{n}\sum\limits_{i=1}^n\log p(\bfx^{(i)}_U | \bfx^{(i)}_{\partial U};\theta')
\end{eqnarray}
for 161 evenly spaced $\theta'$ values ranging from .3 to .5 (granularity .00125). We found the minimizing $\theta'$ to be the true parameter value of .4.

In (b), we consider a single configuration $\bfx$ on $G$, and let $U_1,\ldots,U_n$ be the $n=198$ rows of $G$ with both an upper and lower boundary row. We computed
\begin{eqnarray}
    \frac{1}{n}\sum\limits_{i=1}^n\log p(\bfx_{U_i} | \bfx_{\partial U_i};\theta')
\end{eqnarray}
for the same 161 $\theta'$ values. In this case, the minimizing $\theta'$ to be .4025.

\vspace{2mm}

\section{Discussion}

In this paper we have elaborated on the concept inherent in Maximum Pseudo-Likelihood, namely, that of using conditioning to simplify the task of parameter estimation, and have posed the problem as one of Minimum Conditional Description Length. The specific setting we have considered differs from the typical setting of MPL in that we have in mind temporal rather than spatial invariance, and we have here focused only on estimation of parameters within a single subset. Relaxing the spatial invariance assumption broadens the class of graphs and accompanying parameters to which we can apply this method. However, by requiring temporal stationarity we have imposed a new set of restrictions. More substantively, though, we feel that framing the problem as one of minimizing conditional description length is very natural given that Markov/Gibbs fields are specified by their conditional distributions. This leads to the same MPL estimate when applied to a single configuration generated by a spatially invariant parameter, and as such, we feel that the Minimum Conditional Description Length perspective places the Maximum Pseudo-Likelihood estimate on a firmer theoretical footing.

As we mentioned in the Introduction, though this method can be applied to obtain estimates $\hat\theta^n_{U_1},\ldots,\hat\theta^n_{U_k}$ for the parameters within different subsets, there is potential inconsistency of these estimates for nodes and edges contained within the intersection of these subsets. While resolving this, for example through alternating direction method of multipliers \cite{boyd2011}, remains to be done, we still believe there is value in the notion of taking a large intractable Markov random field and decomposing it into tractable conditional random fields, on which good parameter estimates can be obtained efficiently and in which exact inference and prediction can be performed with respect to these parameters, conditioned on the boundaries of these subsets. Indeed, it was shown in \cite{wainwright2006} that if the MRF is on an intractable graph, such that suboptimal inference and prediction will be performed with respect to whatever parameters are available, then there can be benefits to incorrectly estimating the parameters. In our case, good estimates would be obtained on each tractable conditional random field, and exact inference could be performed with respect to these parameters, but they may not yield a consistent estimate of the global parameter.

Additionally, the MCDL method for parameter estimation introduced in this paper is complementary with our previous work in using cutsets to simplify the processing, in particular the compression, of intractable MRFs \cite{reyes2010}, \cite{reyes2011}, \cite{reyes2016a}. In these works, there is an initial lossless compression of a cutset of sites, followed by either estimation or optimal lossless conditional compression of the remaining sites given the values on the cutset. If a fixed cutset was to be used in one of these algorithms, then one could simply estimate the parameters of the tractable conditional subfields that would be estimated or compressed given the values on their boundaries.

\end{document}